\documentclass[conference]{IEEEtran}
\IEEEoverridecommandlockouts
\usepackage{cite}
\usepackage{amsmath,amssymb,amsfonts}
\usepackage{algorithmic}

\usepackage[dvipdfmx]{graphicx}

\usepackage{url}
\usepackage{hyperref}

\usepackage{multirow}

\usepackage{color}

\usepackage{textcomp}
\usepackage{xcolor}
\begin{document}

\title{Towards Informative Tagging of Code Fragments to Support the Investigation of Code Clones
}

\author{
\IEEEauthorblockN{Daisuke Nishioka}
\IEEEauthorblockA{\textit{Department of Information Systems Design and Data Science} \\
\textit{Grad. Sch. of Natural Sci. \& Tech., Shimane Univ.}, Matsue, Japan \\
n20m107@matsu.shimane-u.ac.jp}
\and
\IEEEauthorblockN{Toshihiro Kamiya}
\IEEEauthorblockA{\textit{Institute of Sci. \& Eng., Shimane Univ.}\\
Matsue, Japan \\
kamiya@cis.shimane-u.ac.jp}
}

\maketitle

\begin{abstract}
Investigating the code fragments of code clones detected by code clone detection tools is a time-consuming task, especially when a large number of reference source files are available.
This paper proposes (i) a method for clustering a clone class, which is detected by code clone detection tools using syntactic similarity, based on topic similarity by considering its code fragments as sequences of words and (ii) a method for assigning short tags to clusters of the clustering result. 
We also report an experiment of applying the proposed method to packages of an open source operating system.
\end{abstract}


\section{Introduction}
Code clone detection is a static analysis technique used to evaluate the maintainability of software products and find duplicate logic before modifying the source code.
In today's software development world, open source software is reused in all kinds of software development, including commercial software products in enterprises\cite{Franch2013}.
Therefore, when developers perform code clone detection on the software products they are developing or maintaining, they need to consider the possibility that code clones exist between their software products and open source products directly or indirectly reused.

There are types of code-clone detection tools: those that detect code clones of the code fragments that are exact matches except for whitespace (type-1), the code fragments that contain renamed identifiers (type-2), the code fragments contain code fragments with some modifications in syntax (type-3), and the code fragments that are equivalent to each other by some semantics (type-4)\cite{Bellon2007}\cite{Roy2009}.

With a code-clone detection tool, developers can detect code clones between the product they are developing and reference (potentially reused) products.
By investigating the code fragments of the detected code clones, they can look for source files that can be used as a reference for development or to find out what variations exist in the code fragments they are working on.
However, the process of investigating code fragments of code clones for development and/or maintenance can be time-consuming when the source files being analyzed contain code clones in many places or when such code fragments are code clones having many code fragments of other files.

In this paper, we propose a method to support such a task of investigating code fragments of code clones by clustering code fragments of code clones to make it easier to find code fragments to be investigated and by tagging code fragments of code clones in the source files to display cluster information in a short format.

\section{Background}

\subsection{Code-Clone Detection}
Various methods and tools have been proposed for detecting code clones, and they are characterized by how source code is represented and compared as a data structure.
A method using line-by-line hash values\cite{Keivanloo2011}. Some methods using the comparison of token sequences, such as Dups\cite{Baker1995}, CCFinder\cite{Kamiya2002}, CCFinderSW\cite{Semura2017}, Göde's\cite{Gode2011}, SourcererCC\cite{Sajnani2016}, iClones\cite{Ishio20172}, and NiCad\cite{Roy2008}\cite{Svajlenko2017}.
Some methods using the similarity of the included identifiers\cite{Marcus2001}\cite{Yuan2011}\cite{Qian2013}\cite{Bauer2014}.
In particular, \cite{Qian2013} presented a method for generating a 10+ words label to summarize a function of the method, including a code fragment of a code clone, by applying NLP (Natural Language Processing) techniques.
The method proposed in this paper also applies NLP techniques to generate a kind of tags for code fragments, but the purpose is to indicate the differences between code fragments rather than to summarize them, and the labels are short, only one word.
Other emerging technologies including \cite{Li2017}, which uses machine learning to determine the similarity of token sequences.

There are tools CloneDR\cite{Baxter1998} and DECKARD\cite{Jiang2007} that represent source code as ASTs and compare their structures.
Some other tools use machine learning to determine the similarity of ASTs\cite{Buch2019}\cite{Zeng2019}\cite{YGao2019}\cite{ZGao2019}.
Some other tools\cite{Komondoor2001}\cite{Liu2006}\cite{Gabel2008}\cite{Jiang2009}\cite{Higo2011}\cite{Wang2017} create PDGs from source code and compare their structures. In particular, \cite{Zhao2018} uses machine learning to determine the similarity between control flow and data flow.

Approaches to determine the similarity of the metrics of a program's structures have also been proposed, such as measuring the input/output of procedures and the number of fields in a class\cite{Mayland1998}\cite{Higo2008}.

Some approaches use a dynamic analysis or an abstract interpretation to determine similarity.
Studies \cite{Jiang2009}\cite{Su2016}\cite{Mathew2020} proposed the methods to determine the equivalence of code fragments by executing code fragments and comparing the execution results.
MeCC\cite{Kim2011} detects code clones by the equivalence of values computed by an abstract interpretation.
Agec\cite{Kamiya2013} attempted to generate a sequence of possible method calls from source code and detect code clones by their similarity.

\subsection{Code-Clone Investigation}

Hereinafter, when some code fragments are equivalent or similar to each other, the set of such code fragments is called a \textit{clone class} (also called a ``clone set'' or a ``clone group'' in some papers).

Various methods have been proposed for software product maintenance, such as supporting developers in identifying code clones or selecting which code fragments to be investigated.

In \cite{Choi2011}, a method was proposed to determine which code clones should be refactored based on the values of code metrics for the code fragments of the code clones.
In \cite{Krishnan2014}, a method was proposed to compare the differences in code fragments of a clone class by representing the code fragments as PDGs to locate the differences in variable names, method arguments, array dimensions, etc., and to determine whether the code fragments can be refactored, by extracting a new method from the code fragments and removing the code fragments.

In \cite{Wang2014}\cite{Sheneamer2020}, a method was proposed to predict whether a clone class should be refactored based on the complexity metrics and editing history of the code fragments of the clone class using decision trees or machine learning.
In \cite{Mondal2015}, a method was proposed to analyze whether the code fragments of a clone class have been modified at the same time in the history of the product's source files and to identify the clone classes that have code fragments that are modified at the same time.

In \cite{Fluri2007}, a method was proposed to extract semantic modifications (such as changes in conditional expressions and method names) from the source code before and after the modifications.
In \cite{Ray2013}, a method was proposed to detect the situation where a bug is created due to semantically inconsistent modifications after reusing source files or code fragments for porting.

There are GUI tools to support the task of investigating the source files that include code clones: Gemini\cite{Ueda2002}(a GUI front-end of CCFinder), AJDT Visualizer\cite{Tairas2006}(an Eclipse plug-in using CloneDR as a clone detection engine), Shinobi\cite{Kawaguchi2009}(a plug-in for Visual Studio IDE), VisCad\cite{Asaduzzaman2011}(a front-end of NiCad), XIAO\cite{Dang2017}(another plug-in for Visual Studio IDE), and Clone Swarm\cite{Bandi2020}(a tool to display clone classes ranked by frequency of modification).

\section{Code-Clone Detection and Investigation with Large Amounts of Reference Source Code}

Developers developing on open source operating systems can clone and refer to the source files of the operating system, libraries, and applications at hand.
As a result of the so-called ``clone-and-own'' approach, open source products tend to contain many identical source files in different versions, which leads to maintainability\cite{Hata2021} and license incompatibility issues\cite{Ishio20171}. 
Commercial software products are not immune to these problems, as 95\% of them contain code derived from open source products\cite{Franch2013}.

When detecting code clones in such a situation where there is a large amount of source code available for reference, dozens of equivalent or similar code fragments may be detected for a single code fragment of a product under development.
When code clones exist in many parts of the source files under analysis, or when one code fragment is a code clone of many other code fragments in many other files, it becomes time-consuming to investigate such code fragments.

\begin{figure*}[t]
\centerline{\includegraphics[width=0.98\textwidth]{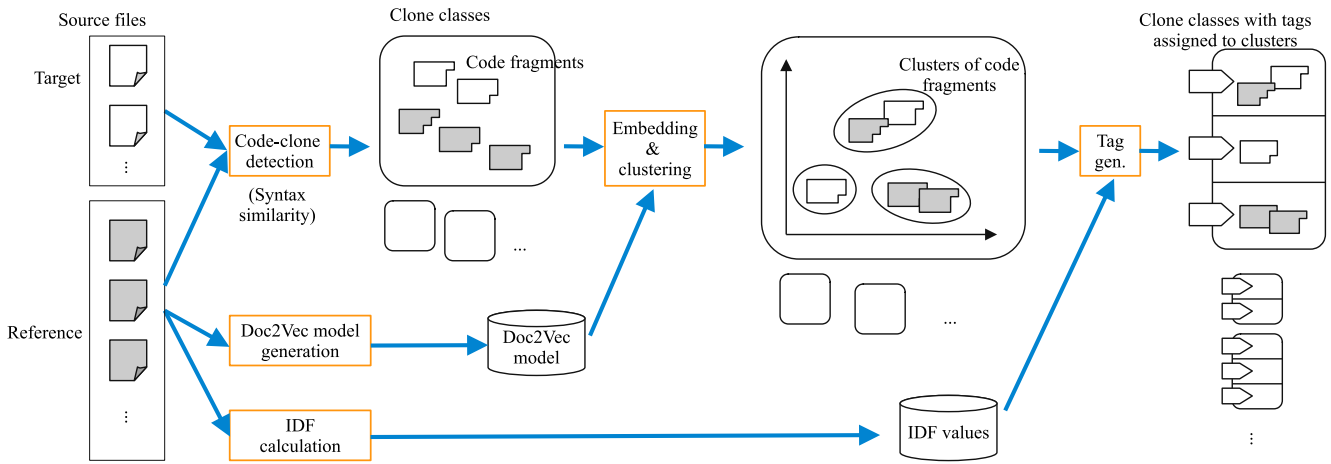}}
\caption{Clustering of Clone Classes and Tag Generation for Clusters.}
\label{fig1}
\end{figure*}

\section{Clustering and Tagging Code Fragments}

The proposed method assumes a situation in which developers apply code clone detection to the (target) product under development and reference products (e.g., open source products that might be potentially reused) and analyze the detected code clones.
The proposed method clusters the code fragments of a clone class and gives each of the resulting clusters a short tag describing the characteristics of the cluster.
Such tags are used to determine which code fragments of the clone class are to be investigated.
For a concrete example of tags, refer to Figure \ref{fig4}, described later in Sec. \ref{subse:tagexample}. 

Figure\ref{fig1} shows the process of generating clusters and tags in the proposed method. 
The proposed method detects code clones, identifies clusters of code fragments for each of the clone classes, and generates appropriate tags (a word or a filename per cluster) to distinguish the clusters.

\subsection{Code-Clone Detection}

As a precondition, the input source files are distinguished into \textit{target} source files and \textit{reference} source files.
Target source files are the source files of the product being developed and maintained.
Reference source files are the source files of libraries, etc., that are potentially reused by the target source files.
The code clone detection extracts the clone classes containing at least one code fragment of any target product.
In other words, each clone class is a set consisting of one or more code fragments of the target product and zero or more code fragments of the reference products.

In the proposed method, we have in mind a code clone detection tool that detects code clones by syntactic similarity.
In the experiment in Sec. \ref{sec:expreiment}, we used CCFinderSW\cite{Semura2017}.

\subsection{Embedding and Clustering}

For each of the clone classes, we cluster the code fragments of the clone class.
In the proposed method, the clustering criterion is the similarity of the numeric vectors that represent the topics of the documents (as defined by Doc2Vec\cite{Le2014}), considering each code fragment as a document made up of a sequence of words.

First, we create a Doc2Vec model from the source files of the reference products.
With this Doc2Vec model, we ``embed'' each of the code fragments into a numerical vector and then cluster these numerical vectors with the k-means method. Euclidean distance is used for k-means clustering. 
The number of clusters is determined dynamically to maximize the silhouette coefficient\footnote{sklearn.metrics.silhouette\_score \url{https://scikit-learn.org/stable/modules/generated/sklearn.metrics.silhouette_score.html}} computed from the clusters.

\begin{figure}[t]
\centerline{\includegraphics[width=0.38\textwidth]{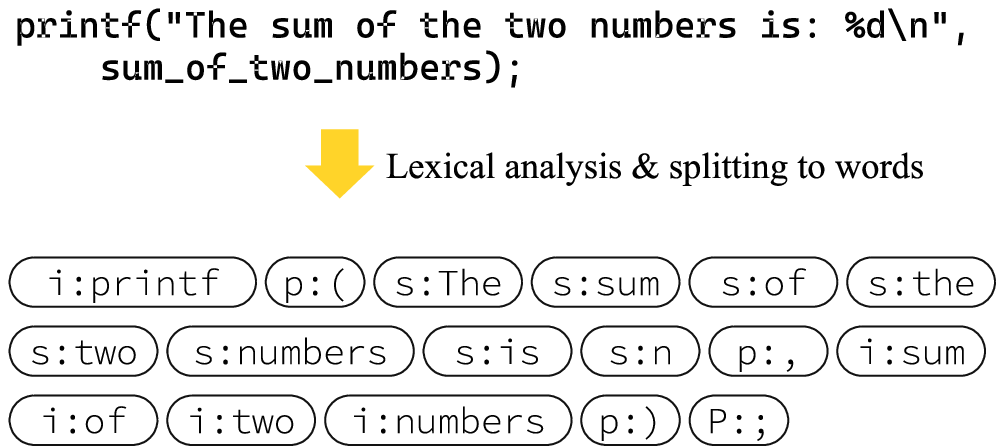}}
\caption{Example of a C code fragment and the word sequence generated from it.}
\label{fig3}
\end{figure}

Doc2Vec is a method for NLP (Natural Language Processing) that embeds documents (expressed as word sequences) into numeric vectors. Numeric vectors become similar when the documents are considered to be dealing with similar topics.
The \textit{words} extracted from the source files are alphabetic words and numbers extracted from identifiers, comments, and literals in the source files, as well as symbols such as delimiters and operators.
For identifiers, we try to extract alphabetic words as much as possible by splitting the identifiers according to camel case and/or underscore. 
A concrete example of extracting alphabetic words from string literals and identifiers is shown in Figure \ref{fig3}.

\subsection{Tag Generation}
\label{sub:taggeneration}

For each clone class, a tag of one word or a filename (filename not including path) is assigned to each cluster the clone class contains.
The reason why tags are as short as one word or a file name is that we assume a UI that displays the source code and shows tags right next to the lines. The tags indicate the number of clusters. Each tag acts as a link to the code fragments of the cluster.

The tag words or filenames are selected based on the following criteria.
\begin{itemize}
\item If filenames of the code fragments in a cluster are all the same, and there is no code fragment of the other clusters with the filename, then the filename is a candidate for a tag of the cluster.
\item A word is a candidate for a tag if it is (i) in the top 3 of the ordered-by-frequency (ObF) word list for all code fragments in the cluster and (ii) not in the top 6 of the ObF word list for code fragments in the other clusters.
\end{itemize}

Here, the \textit{ObF word list} of a given code fragment is defined as a list of words in the code fragment, descending order of their TF-IDF (the number of times they appear in the code fragment multiplied by the inverse document frequency in the reference source files).

\begin{figure}[t]
\centerline{\includegraphics[width=0.49\textwidth]{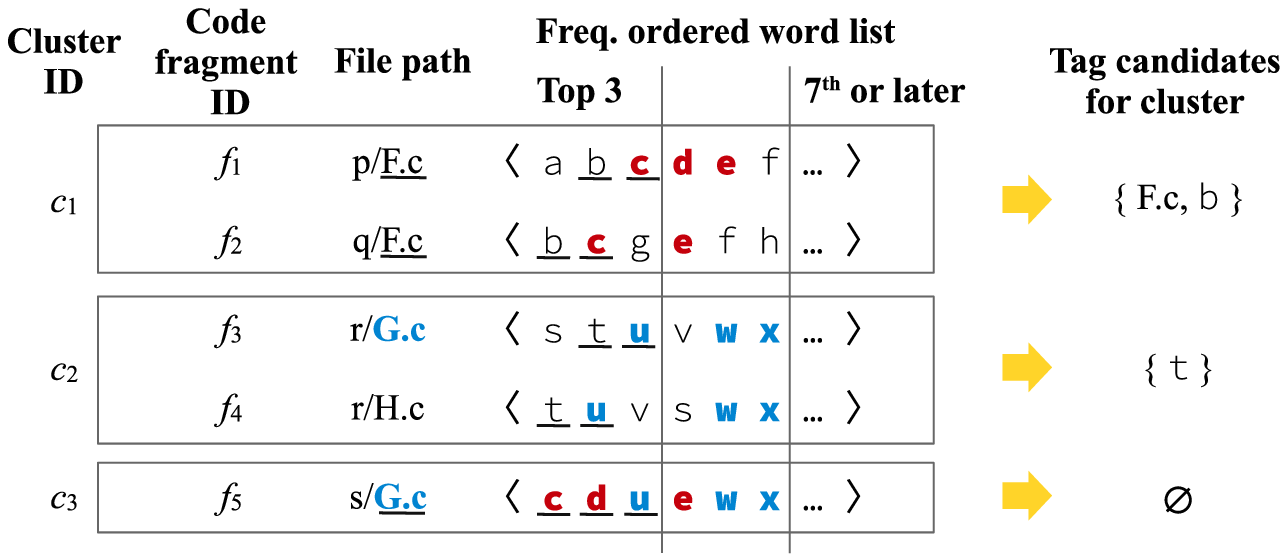}}
\caption{An Illustration of Data about a Clone Class to Explain Tag Generation.}
{\footnotesize Filenames in bold are the filenames of code fragments of two or more clusters. Words in bold appear within the top 6 in the ObF word lists for code fragments of two or more clusters.
An underlined filename is a filename common to all code fragments in a cluster.
An underlined word is in the top 3 in ObF word lists of all code fragments in a cluster.}
\label{fig2}
\end{figure}

\begin{figure*}[t]
\centerline{
\includegraphics[width=0.75\textwidth]{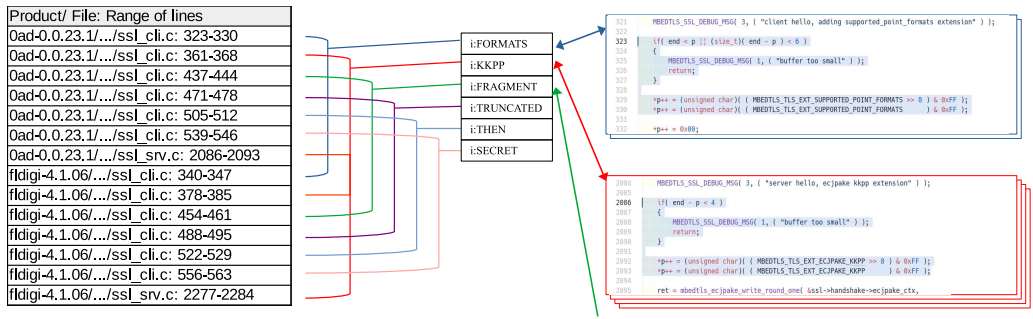}
}
\caption{A Sample of Clusters and Tags in a Clone Class}
\label{fig4}
\end{figure*}
    
The following equation defines the inverse document frequency for a given word:
\[
    IDF(w) = \log\frac{d + 1}{c(w) + 1} + 1
\]
where $w$ is the given word, $d$ is the total number of the reference source files, and $c(w)$ is the number of the reference source files containing the word $w$.

Figure \ref{fig2} illustrates how candidate tags are selected.
For cluster $c_1$, both the filename F.c and the word \verb|b| can be a tag, but the word \verb|c| cannot be a tag because it is included in the top 6 words of cluster $c_3$.
For cluster $c_2$, the word \verb|t| can be a tag, but the word \verb|u| cannot be a tag because it is included in the top 6 words of cluster $c_3$.
For cluster $c_3$, no filename or word can be a tag. The filename G.c exists in cluster $c_2$.
The top 3 words \verb|c|, \verb|d|, and \verb|u| are all included in the top 6 words of cluster $c_1$ or $c_2$.

In the implementation, if a file name could be used as a tag, the file name was preferred over a word.
This was because when comparing words and filenames, the latter was considered to be a better representation of the characteristics of the cluster. This is because the former might represent the meanings of a small part of the file, while the latter is more likely to reflect the meaning of the whole file.

\section{Experiment}
\label{sec:expreiment}

To detect code clones between the target product and each of the reference products, 
the code clone detection tool CCFinderSW\cite{Semura2017} was run by giving each of the reference products together with the target product.
Pairs of the clone classes from different runs were merged when they contained physically identical code fragments (the same range of lines in the same source file of the target product).

For the convenience of the experiment, we developed a tag generation algorithm that generates all possible tags and clustering that satisfy the above conditions for a given clone class,
instead of an algorithm that generates tags for clusters resulting from clustering by Doc2Vec.

In the experiment, we compared the clustering by Doc2Vec with each of the clusterings by tags to see whether we found matching clustering ( that is, whether the clusters by Doc2Vec can be tagged).
In order to evaluate the tag generation method, we also examined whether the clusters could be tagged when only filenames were used as tags.
We set the following two research questions.

\vspace{0.5\baselineskip}
\noindent
\textbf{RQ1} Is it possible for the proposed method to tag all the clusters generated by Doc2Vec?

\vspace{0.5\baselineskip}
\noindent
\textbf{RQ2} Does generating tags only from filenames decrease the number of clone classes that can be tagged?

\subsection{An Example of Tags}\label{subse:tagexample}

The list on the left in Figure \ref{fig4} shows one of the clone classes that was detected. The file name and line number of the code fragment of the clone class are shown.
This clone class has 14 code fragments in 4 source files of 2 products.
In order to investigate the code fragments of this clone class, if there is no hints, you will have to look at these 14 source files in sequence.

In the proposed method, these code fragments were clustered into six clusters according to the similarity of the occurring words.
In addition, for each cluster, a tag was identified. The list in the center of the figure shows the tags.

We assume that a UI will display a code fragment and a list of tags near the code fragment (such as the upper right in the figure). 
The user can see that this code fragment belongs to a cluster with tag \verb|i:FORMATS| (meaning a word ``FORMATS'' in an identifier), 
and that the other clusters having tags other than \verb|i:FORMATS|.
If the user wants to investigate code fragments with different contents as much as possible, he/she can investigate one code fragment for each tag.
When the code in the upper right corner of the figure and the tags are displayed, he/she can investigate the another cluster's code fragments with the expectation that they will have different identifiers from \verb|MBEDTLS_TLS_EXT_SUPPORTED_POINT_FORMATS| of lines 329 and 330.



\subsection{Setup}
The source code used in the experiment was the kernel of the Ubuntu OS and packages such as libraries and tools. They were obtained by the command: \verb|apt source|.
The kernels and packages (hereafter collectively referred to as \textit{products}) that contain C source files (\verb|*.c| and \verb|*.h|) were identified.
We found out that there are multiple kernels such as \verb|linux-5.3.0|, \verb|linux-azure-5.3.0|, \verb|linux-gcp-5.3.0|, etc., so we excluded the directories whose names start with \verb|linux-| other than \verb|linux-5.3.0|.
We used the first package in the lexical order (namely \verb|0ad-0.0.23.1|) as the target product, and the others as the reference products.
Table \ref{tab1} shows the statistics of target and reference products.

A PC equipped with an Intel Corei7-6800K CPU 3.40GHz and 128GiB memory was used for the experiments.
Table \ref{tab3} shows the execution time for each step of the experiment and the amount of peak memory for some of the steps.

Note that the experimental data is available at \url{https://doi.org/10.6084/m9.figshare.14905014.v1}.

\begin{table}[b]
\caption{Statistics of the source code}
\label{tab1}
\begin{center}
\begin{tabular}{|l|r|r|}
\hline
& \textbf{Target} & \textbf{Reference} \\ \hline \hline
Packages (products) & 1 & 11,460 \\ \hline
Files (\verb|*.h|, \verb|*.c|) & 2,059 & 2,063,360 \\ \hline
LOC & 497,426 & 636,407,114 \\ \hline
\end{tabular}
\end{center}
\end{table}

\begin{table}[b]
\caption{Run-time configuration and performance of each step}
\label{tab3}
\begin{center}
\begin{tabular}{|l|l|l|}
\hline
\textbf{Step (Configuration)} & \textbf{Performance (Peak resident memory)} \\ \hline \hline
Clone Detection & 401 min. with 8 processes \\
(-t 50 -tks 12 -rnr 0.3 & (Results for 133 reference projects\\
6,000 reference products) & were not included due to timeout.) \\ \hline \hline
Words extraction & 161 min. with 16 processes \\ \hline
Doc2Vec model generation & 97 min. with 8 threads (8.7GiB) \\ \hline
IDF calculation & 4 min. with single thread (7.2GiB) \\ \hline \hline
Clustering \& Frequent & \multirow{2}*{56 min. with 4 processes} \\
word list generation & \\ \hline
\multirow{3}*{Tag generation} & 44 sec with single thread (101MiB) \\
    & (Ten clone classes were excluded because \\ 
    & the search tree became too large.) \\ \hline
\end{tabular}
\end{center}
\end{table}
    
\begin{table}[b]
\caption{Statistics of the code-clone detection result}
\label{tab2}
\begin{center}
\begin{tabular}{|l|l|r|}
\hline
\textbf{Reference products} & \multicolumn{1}{|r}{5,867} & \\ \hline
\textbf{Clone classes} & \multicolumn{1}{|r}{12,014} & \\ \hline
\textbf{Code fragments (per clone class)} & \multicolumn{2}{|c|}{Max. 355, Min. 2, Ave. 3.81} \\ \hline
\textbf{Clusters (per clone class)$^{\mathrm{a}}$} & \multicolumn{2}{|c|}{Max. 94, Min. 2, Ave. 2.34} \\ \hline
\end{tabular}
\end{center}
{\footnotesize $^{\mathrm{a}}$The values were only for clone classes with two or more clusters.}
\end{table}

\subsection{Clone Detection}

For each of the first 6000 reference products, we ran the code clone detection tool by giving the source files of the product and the target product as input.
We set a timeout of five minutes for each run, and when the run took longer than this, the detection process got stopped\footnote{This experiment was an extreme setting that the code clone detection tool did not anticipate, and therefore a large number of code clones might have been extracted due to the size of the product or the large number of files shared between the two products.
In order to complete the experiment in a reasonable amount of time while running with the parameters assumed by the code clone detection tool, we decided to drop the detection results between some of such pairs of products.}.
In addition, from the detection results of each of these runs, we excluded clone classes that did not contain code fragments of the source files of the target product.
We merged the cloned classes containing even one code fragment of the same (in terms of product, file, or line range). 
The statistics of the clone classes are shown in Table \ref{tab2}. The average code fragment per clone class was 3.81, which means that, on average, two or three code fragments would be candidates when investigating code fragments for a clone class.


\subsection{Words Extraction, Doc2Vec Model generation, and IDF calculation}
\label{subsec:wx}

In order to generate the Doc2Vec model, we converted the source files of the reference products into word sequences.
Each of the documents in the Doc2Vec model is a word sequence generated from each source file.
Using all the reference products was found to take too much time to generate the model, so we used 5\% of the source files from the reference products.
We selected one source file in 20 from the list of source files through all products. 
(The resulting word vocabularies are 432K words taken from identifiers and 404K words in comments.
This size is comparable to about 250K words in the text8 corpus, which is commonly used in NLP processing using Word2Vec.)
To obtain the IDF value for each word, we used the same 5\% of source files.

\subsection{Clustering and Tag Generation}
For each clone class, the code fragments were converted into numerical vectors by Doc2Vec's model and clustering was performed.

Then, we applied a recursive algorithm for each clone class to enumerate all the tags that satisfy the conditions described in Sec. \ref{sub:taggeneration} and the clusterings corresponding to them. 
Such searching for tags got stopped when the number of nodes in the search tree exceeded 100,000.

As an example of clustering by tags, the following clustering C1 using filenames exists for code fragments $f_1$, ..., $f_5$ of the clone class in Figure \ref{fig2}. 
\begin{itemize}
\item[C1] F.c: \{ $f_1$, $f_2$ \}, G.c: \{ $f_3$, $f_5$ \}, H.c: \{ $f_4$ \}
\end{itemize}
Here, each cluster is denoted as ``tag:\{code fragment ID... \}''.
Another clustering by word and filename exists as follows.
\begin{itemize}
\item[C2] F.c: \{ $f_1$, $f_2$ \}, \verb|u|: \{ $f_3$, $f_4$, $f_5$ \}
\end{itemize}
Let's call the clustering in Figure \ref{fig2} as C0, and if we use cluster IDs instead of tags, it can be denoted as follows.
\begin{itemize}
\item[C0] $c_1$: \{ $f_1$, $f_2$ \}, $c_2$: \{ $f_3$, $f_4$ \}, $c_3$: \{ $f_5$ \}
\end{itemize}

By considering a clone class as a set $S$ of code fragments and a clustering as a partition of $S$, we can define a relationship between the clustering:
When any pair of code fragments belonging to the same cluster in a clustering $C$ always belongs to the same cluster in another clustering $D$, we denote it as $C \le D$. When $C \le D$ and $D \le C$ are both satisfied, we denote $C = D$.
When $C \le D$ and not $D \le C$, we denote $C < D$ and say ``$C$ is a refinement of $D$.'' 
The relation between clusterings is a partial ordered relation which, given two clusterings $C$, $D$, can be one of $C = D$, $C < D$, $C > D$, or none of them (for convenience, we denote $C \nsim D$ in the last case).
For example, between C0, C1, and C2 above, the relations C0 $<$ C2, C1 $<$ C2, and C0 $\nsim$ C1 hold.

Table \ref{tab4} shows the results of comparing clusterings by Doc2Vec and ones by tags.
Note that this table only shows the results for the clone classes that resulted in two or more clusters in the Doc2Vec clustering.

\vspace{0.5\baselineskip}
\noindent
\textbf{RQ1}
Table \ref{tab4} shows that 32\% (=1403/4394) of the clusterings were tagged with one word or one filename by the proposed tag generation method.
Also, 44\% (=1951/4394) of the clusterings by tags are equivalent to or refinements of the clusterings by Doc2Vec. In such cases, if we allow two or more words or filenames (e.g., ``a or b'') as a tag for one cluster, we can map multiple clusters by tags to one Doc2Vec cluster, that is, it is possible to tag those Doc2Vec clusterings.
However, the percentage of clusterings that could be tagged does not exceed half (44\% or less) in any case.
For the remaining 56\%, we currently have no choice but to give up on tagging and represent the clusters by serial numbers or metric values.
So further investigation and improvement of the method is needed.

\vspace{0.5\baselineskip}
\noindent
\textbf{RQ2}
Table \ref{tab4} shows that by using words and filenames as tags, we were able to tag more clusterings by Doc2Vec than using filenames solely (1403 vs. 846).
This tendency was also true even under the assumption of using two or more words or filenames as a tag for one cluster (1951 vs. 1494).

\begin{table}[b]
\caption{Comparison of clustering results between Doc2Vec and tags}
\label{tab4}
\begin{center}
\begin{tabular}{|l|r|r|r|r|}
\hline
\textbf{Clustering w/} & $=$ d2vc & $\le$ d2vc & $\ge$ d2vc & $\nsim d2vc$ \\ \hline \hline
Doc2Vec & 4,394 & - & - & - \\ \hline
Words and filenames & 1,403 & 1,951 & 4,207 & 31 \\ \hline
Filenames only & 846 & 1,494 & 3,316 & 430 \\ \hline
\end{tabular}
\end{center}
{\footnotesize The ``Doc2Vec'' row contains the number of clone classes for which the clustering by Doc2Vec resulted in two or more clusters.
The ``$=$ d2vc'' column shows the numbers of clone classes whose clustering by the tag by the left side was the same as clustering by Doc2Vec.
The ``$\le$ d2vc'' column shows the numbers of clone classes whose clustering by the tag by the left side was the same or refinement of clustering by Doc2Vec.
}
\end{table}
    
\section{Threats to Validity}

\subsection{Criteria in Code Clone Detection and Clustering}
In the proposed method, we used CCFinderSW as a code clone detection tool, which determines the similarity of code fragments by their syntax and detects type 1 and 2 code clones.
For clustering of clone classes, we used Doc2Vec, which predicts the topics of code fragments from their containing words.

There might be a better combination of similarity of code clone detection and clustering criteria.
In particular, the effectiveness of the proposed method when applied to type 3 and 4 code clone detection tools has not been evaluated.

\subsection{Generality of the Results of the Experiment}
Because the proposed method was applied to only one target product in the experiment, the characteristics of the target product might have affected the evaluation.
We need to repeat the same experiment by selecting more target products.

\subsection{NLP Issues}
When extracting words, we used camel case and snake case to split identifiers. Also, changes in words due to abbreviations or singular/plural were not taken into account. The proposed method and its implementation did not include reverting abbreviations back to their original English words, but such a process might allow the Doc2Vec model to cluster code fragments by topic more accurately.
Sampling only 5\% of the corpus to calculate the Word2Vec model and IDF values also might have affected the accuracy.

\subsection{Run-Time Performance}
In this experiment, code clone detection, Doc2Vec model generation, IDF values calculation, clustering, and tag generation required hourly computation time. 
However, Doc2Vec model and IDF values do not need to be recalculated unless the reference products change, so there still seems to be potential for speedup.

\subsection{Evaluation in Tasks of Investigating Code Fragments}
The experiment was designed to evaluate whether the proposed tag generation method can select words that can be tags for clusters by Word2Vec.
It serves as a preparation for future evaluation of the human task of analyzing code clones with such tags.

\section{Conclusion}
We proposed a method of tagging clone classes with one word or one filename for clusters clustered by Doc2Vec and k-means+silhouette coefficient.
In the experiments, the proposed method was applied to packages of Ubuntu, an open source OS, and when code clones were detected focusing on a specific package, 32\% of the clone classes could be tagged appropriately. When the limitation was loosened to allow more than two words to be used as a tag, 44\% of the clone classes were tagged appropriately.

As a next step, it is necessary to improve the method so that more clone classes can be appropriately tagged and to evaluate whether the tags generated by the proposed method are useful for investigating code fragments of code clones.

\end{document}